\documentclass[aps,preprint,amssymb,12pt,floatfix]{revtex4}
\usepackage{subfigure}
\usepackage{graphicx}
\usepackage{rotating}

\begin{document}
\title{Pathways and kinetic barriers in mechanical unfolding and refolding of RNA and proteins}
\author{Changbong Hyeon$^1$, Ruxandra I Dima$^3$ \& D. Thirumalai$^{1,2}$}
\thanks{Corresponding author phone: 301-405-4803; fax: 301-314-9404; thirum@glue.umd.edu}
\affiliation{$^1$Biophysics Program, Institute for Physical Science and Technology, University of Maryland, College Park, MD 20742 \\
$^2$Department of Chemistry and Biochemistry, University of Maryland, College Park, MD 20742\\
$^3$Department of Chemistry University of Cincinnati Cincinnati, Ohio 45221\\
\[\]
}
\baselineskip = 22pt
\begin{abstract}
Using self-organized polymer models, we predict mechanical unfolding and refolding pathways of ribo-zymes, and the green fluorescent protein. In agreement with experiments, there are between six and eight unfolding transitions in the Tetrahymena ribozyme. Depending on the loading rate, the number of rips in the force-ramp unfolding of the Azoarcus ribozymes is between two and four. Force-quench refolding of the P4-P6 subdomain of the Tetrahymena ribozyme occurs through a compact intermediate. Subsequent formation of tertiary contacts between helices P5b-P6a and P5a/P5c-P4 leads to the native state. The force-quench refolding pathways agree with ensemble experiments. In the dominant unfolding route, the N-terminal a helix of GFP unravels first, followed by disruption of the N terminus b strand. There is a third intermediate that involves disruption of three other strands. In accord with experiments, the force-quench refolding pathway of GFP is hierarchic, with the rate-limiting step being the closure of the barrel.
\end{abstract}
\maketitle
\newpage

\section{INTRODUCTION}
Despite significant advances (Onuchic and Wolynes, 2004; Thirumalai and Hyeon, 2005), major unsolved problems remain in our understanding of how monomeric RNA and protein molecules navigate the rough energy landscape to reach their folded states. Single-molecule experiments, which use mechanical force to manipulate the initial conformations, have begun to provide a deeper understanding of the folding mechanisms of proteins (Fernandez and Li, 2004; Cecconi et al., 2005) and RNA (Onoa et al., 2003; Liphardt et al., 2001). A combination of forced unfolding and force-quench refolding of a number of proteins (Li et al., 2006; Fernandez and Li, 2004; Cecconi et al., 2005; Best and Hummer, 2005; Isra-lewitz et al., 2001; Gerland et al., 2003) and RNA, including the polynucleotide L-21 \emph{Tetrahymena thermophila} ribozyme (Onoa et al., 2003), and the large (230 amino acid residues) green fluorescent protein (GFP) (Dietz and Rief, 2004) has been used to map the energy landscape of RNA and proteins. These experiments identify kinetic barriers and the nature of intermediates by using mechanical unfolding or refolding trajectories that monitor the end-to-end distance (R(t)) of the molecule in real time (t) or from the force-extension curves (FECs).

It is difficult to unambiguously infer the structural details of the intermediates by using only R(t) and FEC. Moreover, assigning kinetic barriers from FECs or from the distribution of unbinding forces (or rates) is not always unique (Derenyi et al., 2004; Hyeon and Thirumalai, 2006). The power of single-molecule force spectroscopy is enhanced when combined with reliable computations that can be carried out under conditions that mimic the experimental conditions as closely as possible. Toward this end, we use a self-organized polymer (SOP) model (see Experimental Procedures) to predict the forced unfolding and force-quench refolding of the L-21 Tetra-hymena thermophila ribozyme and GFP. Several studies (Chen and Dill, 2000; Treiber and Williamson, 2001; Sosnick and Pan, 2003; Thirumalai et al., 2001; Das et al., 2003) have shown that the RNA energy landscape is rugged. The SOP model is also used to obtain a number of new, to our knowledge, results for mechanical unfolding and force-quench refolding of the large-sized protein GFP whose folding has been difficult to probe by using conventional experiments because the slow folding times often lead to aggregation (Zimmer, 2002). As a result, only a few ensemble folding experiments for GFP (Zimmer, 2002; Fukuda et al., 2000), which is used as a marker in a number of biotechnology applications that include its use as a reporter gene and as a fusion tag to visualize cellular events, have been performed.

Here, we make significant advances in using coarse-grained models to study single-molecule force spectroscopy of large RNA and proteins. The use of the SOP model has enabled us to probe the structural details of the forced-unfolding pathways of the T. thermo-phila ribozyme and related ribozymes and GFP over a wide range of loading rates. For RNA and proteins, the dominant unfolding pathway depends on the loading rate, $r_f$. After establishing the validity of the method by successfully obtaining the experimentally inferred major unfolding pathway for T. thermophila ribozyme, we predict the order of $r_f$-dependent unfolding events in the Azoarcus ribozyme. Application to GFP reveals the structural details of the intermediates identified in forced-unfolding AFM experiments. Refolding simulations upon force quench of the independently folding subdomain P4-P6 of the L-21 construct and GFP show that the assembly of these molecules from stretched states occurs in stages. In both cases, tertiary interactions that stabilize the native conformation form between preformed secondary structural elements. Thus, upon force quench, P4-P6 and GFP refold in a hierarchical manner (Brion and Westhof, 1997; Scalvi et al., 1998; Baldwin and Rose, 1999).
\\

\section{RESULTS and DISCUSSION}

{\bf Summary of the SOP Model: }
Before presenting the results of our work on a variety of systems, it is useful to discuss more fully the advantages and the limitations of the SOP model. In order to predict the pathways in forced unfolding and force-quench re-folding of proteins and RNA under conditions (pulling speeds or loading rates) that are close to those used in experiments, it is necessary to use coarse-grained models (see Experimental Procedures). The effective interactions between the sites in the coarse-grained representation of proteins and RNA involve averaging over degrees of freedom that cannot be easily or fully resolved in experiments. This is the case in laser optical tweezer (LOT) and atomic force microscopy (AFM) experiments, which cannot resolve structures on length scales that are much smaller than about 1 nm. Moreover, in the interpretation of the FECs of RNA and proteins, it is tacitly assumed that unraveling of secondary structures occurs in blocks. In other words, the released length that corresponds to a given number of nucleotides or residues is assigned to the rupture of specific secondary structures in the case of proteins, or hairpins in the case of RNA. With these observations, we constructed the SOP model and kept only one interaction site for each nucleotide or amino acid residue. Such a procedure for coarse graining has already been used in building models for much larger complexes (Sali et al., 2003), where it is not possible (or necessary) to take into account atomic details. The energy function that we chose is simple and consists of terms that are normally employed in more elaborate descriptions. They include chain connectivity and interactions that stabilize the native structures. In the current version, we neglected interactions between residues or nucleotides that are not present in the native structure, i.e., no attractive non-native interactions are allowed. Neglect of nonnative interactions will not affect the FECs qualitatively or quantitatively because, as stated above, in the current experimental setup, only unraveling that occurs as secondary structure blocks (proteins and RNA) is resolved. Furthermore, in the analysis of FEC it is assumed that once a given local secondary structure unravels it remains stretched until the molecule fully extends. To a large extent, the present computations on unfolding of the ribozyme and GFP support such an interpretation.

The simplicity of the SOP model allows us to use pulling speeds that are employed in AFM experiments. As a result, the forces predicted for GFP are in near quantitative agreement with measured values (see below). In the case of RNA, the pulling speeds used in the simulations are about three orders of magnitude greater than in the LOT experiments. Therefore, the predicted unfolding forces are higher. In contrast to our simulations, the pulling speeds used in all-atom molecular dynamics simulations are between six and eight orders of magnitude greater than in AFM experiments and are nearly ten orders of magnitude larger than in LOT experiments. Thus, it is not possible to reproduce the FECs (the experimental observable) to the accuracy reported here by using model force fields employed in all-atom molecular dynamics simulations. The present approach should be viewed as complementary to the more elaborate models that are often used to provide insights into the role that solvent plays in facilitating mechanical unfolding (Isralewitz et al., 2001; Gao et al., 2002).

In contrast to forced unfolding, the neglect of nonna-tive interactions can affect force-quench refolding pathways and timescales. The exclusive use of native interactions minimizes energetic frustration and helps in creating a positive gradient toward the native structure. However, for large molecules (such as GFP) there is a distinct possibility of the molecule being topologically entangled (even when only interactions between contacts in the native state are included) because of the complexity of the native state structures. Hence, our findings (see below) that force-quench refolding leads to pathways consistent with those inferred from ensemble experiments for both GFP and the P4-P6 subdomain are remarkable (see below) and highly nontrivial. It should be stressed that the use of other more computationally demanding models cannot even begin to simulate force-quench refolding. Given the limitations of other computational methods, the insights gained regarding the predictions for force-quench simulations with the SOP model are encouraging.\\

{\bf The L-21 T. thermophila Ribozyme :}
The folding of the L-21 construct of the Tetrahymena ri-bozyme (Figure 1A) and its independently folding subdomains (P4-P6 and P5abc) in various ionic conditions have been extensively investigated (Das et al., 2003; Thirumalai et al., 2001; Treiber and Williamson, 2001). By probing the unfolding characteristics of increasingly larger constructs of the L-21 ribozyme by using LOT experiments, Onoa et al. (2003) were able to associate the force peaks in the FECs to rips (or rupture) of specific substructures. By using this strategy and two other methods, Onoa et al. (2003) have provided an outline of the forced-unfolding pathway of RNA. They assumed that extension by a certain length corresponds to unraveling of the entire helical substructures. With this assumption, the unfolding pathway of ribozymes can be inferred from FECs alone. In the presence of Mg$^{2+}$, the FEC for the L-21 T. thermophila ribozyme has eight peaks. It is difficult to unambiguously assign the specific paired helices that unravel in the absence of the structure of the T. thermophila ribozyme. The number of peaks in the FECs also varies depending on the specific molecule that is being stretched. In addition, there are multiple unfolding routes (Onoa et al., 2003) that may be indicative of heterogeneity in force-induced unfolding.

As a first step in the validation of the SOP model, we computed the FEC for the T. thermophila ribozyme at three loading rates. The Westhof model (The atomic coordinate of the T. thermophila ribozyme, TtLSU.pdb, was obtained from the Group I and II sections in the website http://www-ibmc.u-strasbg.fr/upr9002/westhof.) (Lehnert et al., 1996) was used as the initial conformation in the Brownian dynamics simulations. In agreement with LOT experiments (Onoa et al., 2003), we find that in the majority of unfolding cases the FECs have about eight peaks (Figures 1B-1D). It should be emphasized that the number of peaks varies from molecule to molecule just as in LOT experiments. Such variations from sample to sample are characteristic of single-molecule experiments. By explicitly comparing the FECs and the dynamics of the rupture history (Figures 1C and 1D), we can read off the molecular events leading to the rips. We find two major classes of unfolding pathways. One is [N]$\rightarrow$[P9.2]$\rightarrow$[P9.1, P9, P9.1a]$\rightarrow$[P2]$\rightarrow$[P2.1]$\rightarrow$[P3, P7, P8]$\rightarrow$[P6]$\rightarrow$[P4, P5]$\rightarrow$[P5a, P5b, P5c] (Figure 1C). Helices in square brackets unravel nearly simultaneously. Unfolding can also occur by an alternate route in which the order of rupture is [N]$\rightarrow$[P2]$\rightarrow$[P2.1]$\rightarrow$[P9.2]$\rightarrow$[P9, P9.1, P9.1a]$\rightarrow$[P3, P7, P8]$\rightarrow$[P6]$\rightarrow$[P4, P5]$\rightarrow$[P5a, P5b, P5c] (Figure 1D). The difference between the two pathways is the switch in the order of unfolding of the peripheral domains (P2 and P9). Given the topology of the native structure, this is a significant variation (see the end of the subsection). The experimentally inferred pathway is [N]$\rightarrow$[P9.2]$\rightarrow$[P9.1]$\rightarrow$[P9, P9.1a]$\rightarrow$[P2, P2.1]$\rightarrow$[P3, P7, P8]$\rightarrow$[P6, P4]$\rightarrow$[P5]$\rightarrow$[P5a, P5b, P5c]. The simulation
results for the first pathway are in excellent agreement with most probable pathway inferred from experiments.

The structures in Figure 2 give a visual representation of the conformational changes that occur in the unfolding transition. The advantage of the simulations is that they provide the structural details, albeit at the coarse-grained level, of the populated intermediates and an estimate of their lifetimes. Because of the differences between the loading rates and the spring constant used in the simulations (see caption to Figure 1) and experiments, the predicted FECs do not quantitatively agree with the measurements. However, the order of unfolding of the helices and the heterogeneous nature of the unfolding pathways are in very good accord with experiments.

A few additional comments about our results are worth making. First, both simulations and experiments (Onoa et al., 2003) find that the peripheral domains unravel before disruption of the tertiary interactions involving the catalytic core. Complete rupture occurs when helices P6, P4, and P5abc unfold. Second, the unfolding pathways depend critically on the loading rate, $r_f$. At the lowest loading rate in our simulations, the predicted unfolding pathways coincide with the results of Onoa et al. (2003). As $r_f$ increases by an additional factor of four, we find that the catalytic core (P3-P7-P8) interactions unravel before P2 (data not shown). At even higher loading rates, the T. thermophila ribozyme unfolds sequentially,
starting from the P9 domain and ending at the P2 domain. The order of unfolding is determined by $r_f$ and by the rate at which tension propagates along the structure. Third, it is worth stressing that the two unfolding pathways are not trivially related to each other. In one pathway, unfolding is initiated from P2, while in the other unraveling starts from the P9 end. From a structural perspective, P2 forms tertiary interactions with P5c (Figure 1A), whereas the P9 helix is in contact with P5. The free energies of the tertiary interactions involving the P2 and P9 domains are also different. Thus, from both the energetic and structural considerations, the differences in the unfolding pathways are significantly different. Accurate prediction of the pathways and the associated $r_f$-dependent amplitudes requires a combination of simulations and experiments. Fourth, the rips corresponding to the peripheral domains P9 in the simulations are [P9.2]$\rightarrow$[P9.1, P9, P9.1a], whereas in the experiments three rips corresponding to [P9.2]$\rightarrow$[P9.1]$\rightarrow$[P9, P9.1a] are identified. The two rips corresponding to [P2]$\rightarrow$[P2.1] from the peripheral domains P2 also varies from the single rip [P2, P2.1] in the experiment. The minor differences are due to the slight variations in the constructs used in experiments and simulations. The LOT experiments used the L-21 construct that contains 390 nucleotides whose secondary structure map shows (Onoa et al., 2003) that P1 is not present. The T. thermophila ribozyme used in the simulations is longer (407 nucleotides) and, in addition to the presence of the P1 helix, has a slightly longer extension at the 3' end.
\\

{\bf Forced-Unfolding Pathways of the Azoarcus Ribozyme Depend on the Loading Rate:}
An important prediction of the SOP model for the T. thermophila ribozyme is that the very nature of the unfolding pathways can drastically change depending on $r_f$. This suggests that outcomes of unfolding by LOT and AFM experiments can be different. In addition, predictions of forced unfolding based on all-atom MD simulations should also be treated with caution unless, for topological reasons (as in the Ig27 domain from muscle protein titin) (Isralewitz et al., 2001; Klimov and Thirumalai, 2000a; R.I.D. and D.T., unpublished data), the unfolding pathways are robust to large variations in the loading rates. In order to fully explore the origins of the changes in the unfolding pathways as $r_f$ is varied, we have simulated the rip dynamics of the Azoarcus ribozyme (Figure 3A) for which experimental data are not yet available. The structure of the smaller (195 nt) Azoarcus ribozyme (Rangan et al., 2003) (PDB code: 1u6b) is similar to the catalytic core of the T. thermophila ribozyme including the presence of a pseudoknot. The reduction in the number of nucleotides allows us to explore forced unfolding over a wide range of loading conditions. For the Azoar-cus ribozyme, we generated ten mechanical unfolding
trajectories at three loading rates. At the highest loading rate ($r_f\approx 2.4\times 10^5 r_f^{LOT}$), the FEC has six conspicuous rips (red FEC in Figure 3B), whereas at the lower $r_f$ the number of peaks is reduced to between two and four. We identify the structures in each rip by comparing the FECs (Figure 3B) with the history of rupture of contacts (Figure 3C). At the highest loading rate, the dominant unfolding pathway of the Azoarcus ribozyme is N$\rightarrow$[P5]$\rightarrow$[P6]$\rightarrow$[P2]$\rightarrow$[P4]$\rightarrow$[P3]$\rightarrow$[P1]. At medium loading rates, the ribozyme unfolds via N$\rightarrow$[P1, P5, P6]$\rightarrow$[P2]$\rightarrow$[P4]$\rightarrow$[P3], which leads to four rips in the FECs. At the lowest loading rate, the number of rips is further reduced to two, which we identify with N$\rightarrow$[P1, P2, P5, P6]$\rightarrow$[P3, P4]. Unambiguously identifying the underlying pulling speed-dependent conformational changes requires not only the FECs, but also the history of rupture of contacts (Figure 3C).

To understand the profound changes in the unfolding pathways as $r_f$ is varied, it is necessary to compare $r_f$ with $r_T$, the rate at which the applied force propagates along RNA (or proteins). In both AFM and LOT experiments, force is applied to one end of the chain (30 end) while the other end is fixed. The initially applied tension propagates over time in a nonuniform fashion through a network of interactions that stabilize the native conformation. The variable $\lambda = r_T/r_f$ determines the rupture history of the biomolecules. If $\lambda \gg 1$, then the applied tension at the 50 end of the RNA propagates rapidly so that, even prior to the realization of the first rip, force along the chain is uniform. This situation pertains to the LOT experiments (low $r_f$). In the opposite limit, $\lambda\ll 1$, the force is nonuniformly felt along the chain. In such a situation, unraveling of RNA begins in regions in which the value of local force exceeds the tertiary interactions. Such an event occurs close to the end at which the force is applied.
The intuitive arguments given above are made precise by computing the rate of propagation of force along the Azoarcus ribozyme. To visualize the propagation of force, we computed the dynamics of alignment of the angles between the bond segment vector ($r_{i, i + 1}$) and the force direction during the unfolding process (Figures 3D \& 3F). The nonuniformity in the local segmental alignment along the force direction, which results in a heterogeneous distribution of times in which segment vectors approximately align along the force direction, is most evident at the highest loading rate (Figure 3E). Interestingly, the dynamics of the force propagation occurs sequentially from one end of the chain to the other at high $r_f$. Direct comparison of the differences in the alignment dynamics between the first ($\theta_1$) and last angles ($\theta_{N-1}$) (see Figure 3D) illustrates the discrepancy in the force values between the 30 and 50 ends (Figure 3F). There is nonuniformity in the force values at the highest $r_f$, whereas there is a more homogeneous alignment at low $r_f$. The microscopic variations in the dynamics of tension propagation are reflected in the rupture kinetics of tertiary contacts (Figure 3C) and, hence, in the dynamics of the rips (Figure 3B).\\

{\bf Force-Quench Refolding of the P4-P6 Domain:}
Folding of the T. thermophila ribozyme induced by increasing counterions is complicated because of pausing in kinetic traps. In order to dissect the folding pathways of the larger ribozyme the independently folding P4-P6 domain has often been studied (Laederach et al., 2006; Uchida et al., 2002; Deras et al., 2000). Because of the interplay of a number of distinct timescales, it has not yet been possible to unambiguously produce the refolding pathways by using bulk experiments alone. In principle, mechanical force can be used as a way to initiate folding, as was shown in the context of protein folding (Fernandez and Li, 2004). We have followed the AFM experimental procedure to monitor refolding by quenching, $f$, from a high to a low ($f_Q$) value. We used force-quench simulations to predict the refolding dynamics of P4-P6, starting from an initially stretched state. By setting $f_Q = 0$, we monitor the dynamics of the transition from a low-entropy (rod-like initial state) structure to a folded low-entropy final state. We generated 20 mechanical refolding trajectories (see Figure 4A for an example) for the P4-P6 domain (the domains enclosed by the blue rectangle in Figure 1A) by using the coordinates of the crystal structure (PDB code: 1gid) in the simulations. The time dependence of formation of the fraction of native contacts, $Q(t)$, shows that the approach to the native state conformations occurs in steps (Figure 4A). After an initial rapid increase in Q(t) in $t\approx2 ms$, there is a long pause (ranging from 2 to 15 ms) in a metastable state with $Q(t)\sim 0.65$. A rapid cooperative transition to the folded state occurs at $t =\sim 6 ms$, with $Q(t \approx 6 ms)\approx0.8$. The ends of the chain, monitored with $R(t)$, reach the native value rapidly (Figure 4B).
The compaction of the P4-P6 domain that is monitored by using the radius of gyration ($R_G(t)$) decreases in two distinct stages. After a rapid initial collapse that results in a sharp reduction of $R_G(t)$ by about 10 nm, a more gradual approach to the native value ($R^N_G=3.0 nm$) occurs (Figure 4B). The time dependence of the root mean square deviation, $\Delta(t)$, decreases in stages and abruptly drops around $t\approx 6 ms$ from $\sim 3 nm$ to $\sim 0.7 nm$ (Figure 4B). The two-stage compaction of RNA appears to be a robust mechanism of chain collapse. Indeed, ensemble experiments that have probed the equilibrium changes in RG of the P4-P6 domain by using small-angle X-ray scattering experiments also show a two-stage relaxation to the native state (Takamoto et al., 2004). As the counterion concentration increases, the value of RG decreases without forming tertiary interaction. At a higher ion concentration, the native state is reached with the formation of tertiary interactions. The dynamic processes shown in Figure 4B mirror the equilibrium pathway inferred from SAXS experiments. In contrast to folding initiated by increasing the counterion concentration, much larger changes in RG occur in the early stages when force quench begins from a fully stretched state.

Examination of the tertiary contact formation at the nucleotide level (Figure 4C) shows that the assembly of the native structure is initiated from the P5b and P6a hairpin loop within about 5 ms. Subsequently, zipping of the secondary structures takes place (the structure in Figure 4A). Formation of loops from the initial single-stranded structure is the key nucleation event that triggers formation of the helices. The rate-determining step in the native state formation involves a search in the ensemble of conformations with preformed secondary structures. Upon cooperative formation of the tertiary contacts involving P5b-P6a and P5a/P5c-P4 (Figure 4C), structural transition to the native state occurs. The force-quench refolding mechanism reveals the hierarchical nature of RNA structures (Brion and Westhof, 1997; Scalvi et al., 1998).
It has been difficult to precisely pinpoint the refolding of the various structural elements of the P4-P6 domain by using ensemble experiments (Uchida et al., 2002; Laederach et al., 2006). However, a wealth of experimental data and recent novel analysis methods (Laederach et al., 2006) reveal that the early formation of hairpins in the P5abc domain (nearly simultaneous formation of P5c and P5b) directs the refolding of the P4-P6 domain of the Tetrahymena ribozyme. These events are followed by tertiary interactions between P5b-P6a and the helices from the P5abc subdomain and the P4 and P6 helices. Remarkably, the inferred pathway from ensemble experiments (Laederach et al., 2006; Uchida et al., 2002) is almost coincident with the force-quench simulations presented here. The limited comparison of the relaxation of RG and the approach to the native state obtained by using simulations and experiments shows that the SOP model is successful in identifying refolding pathways of complex RNA structures. It would be interesting to experimentally validate the predictions by performing force-quench experiments at the single-molecule level.\\

\section{Application to Proteins}
The versatility of the SOP model is demonstrated by making several nontrivial predictions for forced unfolding and force-quench refolding of GFP that has a complex three-dimensional structure (Figure 5A). We undertook these simulations to provide insights into constant loading rate AFM experiments on GFP that were used to construct its partial energy landscape (Dietz and Rief, 2004). In the AFM experiments, two unfolding intermediates were identified. Disruption of H1 (Figure 5A) results in the first intermediate, GFPDa. The second intermediate, GFPDaDb, was conjectured to be either unraveling of $\beta1$ from the N terminus or $\beta11$ from the C terminus (Figure 5A). Both $\beta1$ and $\beta11$ have the same number of residues, making the assignment of the strand that unravels first by using FEC alone impossible (Dietz and Rief, 2004). In general, precise assignment of the structural characteristics of the intermediates with FEC alone is difficult not only because of the complex topology of GFP, but also because, unlike in RNA, the substructures of GFP may be unstable. More generally, because secondary structures in proteins are typically unstable in the absence of tertiary interactions, it is impossible to obtain the unfolding pathways from FEC alone. The native state of GFP (PDB code: 1gfl; Figure 5A) consists of 11 $\beta$ strands, 3 helices, and 2 relatively long loops. A two-dimensional connectivity map of the $\beta$ strands shows that $\beta4$, $\beta5$, $\beta6$ and $\beta7$, $\beta8$, $\beta9$ are essentially disjointed from the rest of the structure (Figure 5B). From the structure alone we expect that the strands in the substructures ($D\beta_1\equiv[\beta4, \beta5, \beta6]$) and ($D\beta_2\equiv[\beta7, \beta8, \beta9]$) would unravel almost synchronously. However, it is not possible to predict the order of unfolding, the diversity of the unfolding pathways, or the number of intermediates without reliable computations with pulling speeds that match those used in the AFM experiments (Dietz and Rief, 2004).
We probed the structural changes that accompany the forced unfolding of GFP by using FECs and the dynamics of rupture of contacts at $v = 2.5 \mu m/s$ ($\sim 2.5 v_{AFM}$). The FECs in a majority of molecules have several peaks (Figure 5C) that represent unfolding of specific secondary structural elements. By following the dynamics of residue-dependent breakup of contacts (Figure 5D), the structures that unravel can be unambiguously assigned to the FEC peaks. Unfolding begins with the rupture of H1 (leading to the intermediate GFP$\Delta\alpha$), which results in the extension by about $\Delta z\approx 3.2 nm$ (Figure 5C). The force required to disrupt H1 is about 50 pN (Figure 5C), which compares well with the experimental estimate of z35 pN (Dietz and Rief, 2004) at the lower pulling speed. In the second intermediate, GFP$\Delta\alpha\Delta\beta$, $\beta1$ unfolds (Dietz and Rief, 2004). The value of the force required to unfold $\beta1$ is about 100 pN (Figure 5C), which is also roughly in accord with the experiment (Dietz and Rief, 2004). The measured unfolding force that corresponds to the second intermediate ranges from 70 to 100 pN (Dietz and Rief, 2004). After the initial events, the unfolding process is complex. For example, ruptured interactions between strands $\beta2$ and $\beta3$ transiently reform (Figures 5C and 5D). The last two rips represent unraveling of D$\beta_1$ and D$\beta_2$, in which the strands in D$\beta_1$ and D$\beta_2$ unwind nearly simultaneously. The structures that remain after the various rips (labeled $[i]-[iv]$ in Figure 5D) are shown in Figure 5E.

Besides the dominant pathway, there is a parallel unfolding route in some of the trajectories. In the alternate pathways (Figure 6A), the C terminus strand $\beta11$ unfolds after the formation of GFP$\Delta\alpha$ (Figure 6B). In both the dominant and the subdominant routes, the simulations identify multiple intermediates. To assess if the intermediates in the dominant pathway are too unstable to be detected experimentally, we have calculated the accessible surface area of the substructures by using the PDB coordinates for GFP. The structures of the intermediates are assumed to be the same upon rupture of the secondary structural elements, and hence our estimate of surface area is a lower bound. The percentage of exposed hydrophobic residues in the intermediate $[\beta2, \beta3, \beta11]$ is 25\%, compared to 17.4\% for the native fold, whereas in excess of 60\% of the hydrophobic residues in $\Delta D\beta_2$ are solvent accessible. We conclude that the intermediate $[\beta2, \beta3, \beta11]$, in which $H1$, $\beta1-\beta3$, and $\beta11$ partially unfolds, is stable enough to be detected. However, the lifetimes of the late-stage intermediates are likely to be too short for experimental detection. In the subdominant unfolding route, the barrel flattens after the rupture of $\beta11$, thus exposing in excess of 50\% of hydrophobic residues. As a result, we predict that there are only two detectable intermediates.

{\bf Refolding of GFP upon Force Quench}
To initiate refolding, we reduced the force on the fully stretched GFP to $f_Q = 0$. Formation of secondary structures and establishment of a large number of tertiary contacts occurs rapidly in w0.25 ms (Figure 7). Subsequently, the molecule pauses in a metastable intermediate state with $Q(t)\approx0.7$ in which all of the secondary structural elements are formed but the characteristic barrel of the native state is absent. The transition from the metastable intermediate to the native basin of attraction, during which the barrel forms, is the rate-limiting step that occurs abruptly, with $Q(t)$ reaching 0.8 (Figure 7A). Native state formation is signaled by the closure of the barrel and the accumulation of the long-range contacts between H1 and the rest of the structure. Both $R(t)$ and $R_G(t)$ decrease nearly continuously, and only in the final stages is there a precipitous drop in $R(t)$ and $\Delta(t)$ (see inset in Figure 7A). The time dependence of $\Delta(t)$ shows that the root mean square deviation of the intermediate from the native state is about 20 \AA, whereas the final refolded structure deviates by only $\approx$ 3 \AA\ from the native conformation. Contact formation at the residue level (Figure 7B) shows that interactions between $\beta3$ and $\beta11$ and between $\beta1$ and $\beta6$ are responsible for barrel closing. The assembly of GFP appears to be hierarchical in the sense that the secondary structural elements form prior to the establishment of the tertiary interactions.

It is interesting to compare the force-quench refolding results to known pathways that have been inferred from kinetics of Cycle 3 GFP refolding from an acid-denatured state by using stopped-flow CD and fluorescence techniques (Enoki et al., 2004; Fukuda et al., 2000). Because the chemical structure of the p-hydroxybenzyli-deneimidazolidone chromophore remains intact in the acid-denatured state, they used the green fluorescence of the chromophore to monitor the formation of the native structure. Within the dead time of the instrument they observed a nonspecific collapse that in our simulations is manifested as a sharp decrease in RG (see the inset of Figure 7A). They proposed that there is partial secondary structure formation in the nonspecific collapse state. Such an interpretation was made in light of the increase in tryptophan fluorescence that showed that Trp57 (the only tryptophan in GFP) is solvent inaccessible. This implies that there is at least partial ordering of the structure around position 57. On the other hand, the absence of chromophore fluorescence in this phase indicates that the chromophore (which lies more toward the C-terminal end of the chain than Trp57) is still solvent accessible due to a lack of formation of a rigid specific structure around it. In good agreement with these results, in our simulations we found that the first native contacts to form are those involving residues 1-65 (i.e., spanning the $\beta1-\beta3$ region), which are in the neighborhood of Trp57 (see Figure 7B).

The lag phase in the chromophore fluorescence was used to suggest the presence of an on-pathway intermediate that is more compact than the burst-phase intermediate. More importantly, in the intermediate many of the secondary structural elements of GFP (i.e., all of the $\beta$ strands) are formed. However, the barrel is not yet present, as there is very little fluorescence from GFP (i.e., the chromophore is solvent accessible). It is likely that in this intermediate Trp57 is closer in space to the chromophore because its fluorescent signal is quenched. The long-lived metastable intermediate found during our GFP refolding simulations (see the structure in Figure 7A) has the same characteristics as the structure inferred from experiments. Our force-quench simulations show that all of the b strands are formed; therefore, Trp57 is close to the chromophore location. In addition, the characteristic barrel of GFP is not formed. Indeed, we predict the closing of the barrel to be the rate-determining step in the refolding. The explicit comparisons between simulation results and conclusions from bulk experiments that are based on interpretations of spectroscopic data show, just as for RNA, that, even for systems with complex architecture, refold-ing pathways can be reliably predicted using the SOP model.\\

\section{Concluding Remarks}
The ability to monitor folding and unfolding of biopoly-mers, starting from arbitrary regions of the energy landscape, makes force spectroscopy unique. In recent years, LOT and AFM experiments have been used to produce FECs for large RNA and for proteins from which their energy landscapes have been constructed. In order to provide a structural interpretation of the FEC results, it is necessary to use simulations of mechanical unfolding of ribozymes and GFP under loading conditions that are typically used in experiments. We have introduced the SOP model to study mechanical unfolding and force-quench refolding of large RNA molecules and proteins. After establishing the reliability of the model, we have made a number of testable predictions for both RNA and proteins by using the SOP model. For RNA and GFP, we find that the unfolding pathways are critically dependent on the pulling speed. The order of unfolding is determined by the ratio of the applied loading rate to the rate (a topology-dependent variable) at which force propagates along the molecules. Typically, at very high loading rates, unfolding occurs along a unique topology-dependent pathway, whereas as at lower $r_f$ excursions from the dominant pathway can be found. The crucial role played by the loading rate suggests that a detailed picture of the energy landscape can only be obtained by using a combination of LOT and AFM experiments.
The form of the energy function used in the simulations is identical for both RNA and proteins. Despite the simplicity, the model reproduces the experimentally inferred order of unfolding in both RNA and proteins. In the case of GFP, the simulations clearly resolve the nature of the second intermediate and further predict that a third intermediate must be observable in the FEC. In addition, the SOP gives the structures of the intermediates that are almost impossible to obtain from experiments. Because the pulling speeds used in the GFP simulations are similar to those used in experiments, the predicted unbinding forces are in close agreement with measured values. The applications to ri-bozymes and enzymes show that, for certain problems, a unified perspective of the energy landscape governing folding of RNA and proteins can be obtained (Thirumalai and Hyeon, 2005).
The SOP model was used to monitor the dynamics of refolding upon force quench. These simulations are important in interpreting, at the molecular level, experiments that can only obtain the dynamics of the end-to-end distance ($R(t)$) relaxation upon force quench. Our simulations, which identify structural details of the intermediates in the refolding of P4-P6 and GFP, clearly show that the routes explored in the mechanical unfolding process do not coincide with those in the refolding process. The SOP simulations have also been used to obtain the dynamics of that compaction process under folding conditions ($f_Q = 0$) that are extremely difficult to obtain from single-molecule experiments. Even time-resolved small-angle X-ray scattering experiments cannot currently resolve the behavior of RG at short time periods.
The refolding of the P4-P6 subdomain and GFP occurs by a hierarchical mechanism (Baldwin and Rose, 1999; Brion and Westhof, 1997). In RNA, the separation of energy scales involving secondary and tertiary interactions is the reason for the hierarchical assembly. The force-quench refolding of GFP suggests that large
proteins are more likely to follow hierarchical assembly than small, globular proteins. Perhaps, the hierarchical assembly mechanism restricts the severity of pausing for long periods of time in kinetic traps that, in the absence of energetic frustration, arise due to topological frustration (Thirumalai and Woodson, 1996). Future force-quench experiments, at the single-molecule level, would be invaluable in checking the predictions of this work.\\

\section{Experimental Procedures}
{\bf Self-Organized Polymer Model}
We introduce a versatile coarse-grained structure-based model, referred to as a self-organized polymer (SOP) model, that can be adopted to describe forced unfolding of proteins and RNA. In order to simulate force-ramp and force-quench folding and unfolding of large RNA and proteins, under conditions that are close to those used in experiments, we are forced to employ coarse-grained, structure-based models, such as the SOP model (see below), that can be adopted to describe forced unfolding of proteins and RNA. Use of the SOP model to explore mechanical unfolding is justified for the following reasons. First, force-induced unfolding results in a 10-100 nm increase in the end-to-end distance (R). With the current spatial resolution in single-molecule forced-unfolding experiments, it is not possible to resolve the changes at the atomic level, i.e, it is not possible to resolve structural changes on length scales less than 1 nm. As a result, changes at small length scales are masked in mechanical unfolding experiments that only provide direct information on force-extension curves (FECs). We have, therefore, sought models that can reproduce experimental FECs as accurately as possible under loading conditions that are similar to those used in LOT and AFM experiments. Thus, while these models cannot describe changes at the atomic level (Pabon and Amzel, 2006; Gao et al., 2002), they are versatile enough to explore mechanical unfolding and force-quench refolding over a wide range of external conditions. In some sense, the present simulations should be viewed as a complement to the steered molecular dynamics simulations (Isra-lewitz et al., 2001). Second, our previous work (Klimov and Thiruma-lai, 2000a) established that accurate estimates of unfolding forces and the pulling speed variations of the most probable values of the disruption forces for proteins can be made by using the native topology alone. It follows, therefore, that as long as interactions that stabilize the native fold are adequately taken into account, many aspects of forced unfolding can be semiquantitatively predicted.

With the above-described observations in mind, we introduce the SOP model that is suitable for accurately predicting the kinetic barrier in large RNA and proteins. We represent each nucleotide or a residue by using a single interaction center. The total energy function of a conformation, specified in terms of the coordinates {ri} (i = 1, 2, $\ldots$, N), where N is the number of nucleotides or residues, is given by:
\begin{eqnarray}
V_T&=&V_{FENE}+V_{NB}^{ATT}+V_{NB}^{REP}\nonumber\\
&=&-\sum_{i=1}^{N-1}\frac{k}{2}R_0^2\log({1-\frac{(r_{i,i+1}-r_{i,i+1}^o)^2}{R_0^2}})\nonumber\\
&+&\sum_{i=1}^{N-3}\sum_{j=i+3}^N\epsilon_h[(\frac{r^o_{ij}}{r_{ij}})^{12}-2(\frac{r^o_{ij}}{r_{ij}})^6]\Delta_{ij}\nonumber\\
&+&\sum_{i=1}^{N-2}\sum_{j=i+2}^N\epsilon_l(\frac{\sigma}{r_{ij}})^6(1-\Delta_{ij}).
\label{eqn:VT}
\end{eqnarray}
Finite extensible nonlinear elastic (FENE) potential describes the backbone chain connectivity. The distance between two neighboring interaction sites,  $i$ and $i+1$ is $r_{i,i+1}$ and $r_{i,i+1}^o$ is its value in the native structure. We use the Lennard-Jones potential to account for the interactions that stabilize the native state. If the noncovalently linked beads i and j for $|i- j| > 2$ are within a cut-off distance, RC (i.e., $r_{ij} < R_C$), then $\Delta_{ij} = 1$. If $r_{ij} > R_C$, then $\Delta_{ij} = 0$. A uniform value for 3h, which specifies the strength of the nonbonded interactions, is assumed. All nonnative interactions (third term in Equation (1) are repulsive. To impose a constraint on the bond angle between i, i + 1, and i + 2 beads, the repulsive potential is used with parameters determining the strength, $\epsilon_h$, and the range of repulsion, $\sigma_{i, i + 2}$. We used $\sigma_{i, i + 2}= \sigma/2$ for RNA and $\sigma_{i, i + 2}= \sigma$ for proteins. To prevent interchain crossing, we chose the appropriate value of $\sigma$ (see Table 1).

There are six parameters in the energy function (see Table 1 for their values for proteins and RNA). Of these, the results are insensitive to the exact choices of $R_0$ and $k$ that account for chain connectivity. Similarly, the value of $\epsilon_h$, which is introduced to emphasize native interactions and prevent chain overlap, is not critical. The native structure is sensitive to $\epsilon_h$ (more precisely, the ratio $\epsilon_h/\epsilon_l$) and $R_C$. The range of values for $R_C$ is, to a large extent, dictated by the structures of RNA and proteins in the Protein Data Bank (PDB). In the SOP model, the parameter $\epsilon_h$ plays a central role.

A few comments about the model are needed. (1) The model differs from the off-lattice Go model (Clementi et al., 2000; Klimov and Thirumalai, 2000a) in two respects. First, chain connectivity potential in our model is nonlinear, whereas the Go model uses a harmonic potential. We chose the FENE potential because the FECs for RNA and proteins exhibit pronounced curvature at large forces, i.e., even after substantial loss of structure. The nonlinear FENE potential can reproduce the observed curvature in the FEC. Second, we neglect the dihedral angle degrees of freedom that describe local secondary structural preferences. Because of limitations in spatial resolution in LOT or AFM experiments, the disruption of secondary structures cannot be probed. Therefore, we do not include torsional potentials in Equation 1. (2) We used a repulsive potential between nonnative interactions (third term in Equation 1), whose range is different from the normally used inverse 12th potential. In our previous study (Klimov and Thirumalai, 2000b) of $\beta$ hairpin formation, we had used a similar power to approximately mimic hydration effects. In the present context, as long as the range of the repulsive potential is short ranged, the precise power used is not relevant. We chose the inverse sixth power for purely practical purposes. The force-induced Brownian dynamics simulations are performed under nonequilibrium conditions. Under these conditions, we find that it is important that appropriate values of $\sigma$ and $\epsilon_l$ (third term in Equation 1) be chosen to prevent RNA and proteins from nonphysical chain crossing. Furthermore, the simulations have to be performed for long times to observe global unfolding of the large ribozymes and proteins, especially at moderate and low pulling speeds. The need to follow the dynamics for long time periods requires us to choose a relatively large time step. For these practical reasons, we find that the longer-range repulsive potential (inverse sixth power), rather than the harsher shorter-range interactions (inverse 12th power), is more appropriate. Other forms of repulsive potential will not alter the results in any significant way. (3) The major limitation, especially in RNA applications, is that environmental changes (especially the role of counterions) are not adequately taken into account. These effects can be included in an approximate way by varying $\epsilon_h$ (or $\epsilon_l$).\\

{\bf Simulations:} 
Using the SOP model, we simulated the mechanical unfolding of proteins and RNA by using the Brownian dynamics algorithm (Ermack and McCammon, 1978; Klimov et al., 1998), for which the characteristic time for the overdamped motion is $\tau_H = (\zeta\epsilon_hh/k_BT)\tau_L$, where $\tau_L= 4 ps$ for RNA and $\tau_L = 3 ps$ for proteins (Veitshans et al., 1997). After the experimental setup in AFM measurements, the C-terminal end is stretched at a constant pulling speed, while the N-terminal
end is kept fixed. We chose $k_s = 35 pN/nm$, which is in the range of
$1-100 pN/nm$ used in the AFM experiments.
For RNA, since a typical value for the mass of a nucleotide, m, is $\sim 300-400$ $g/mol$; the average distance between the adjacent beads, $a$, is 5 \AA; the energy scale, $\epsilon$, is 0.7 kcal/mol, then the characteristic time is $\tau_L = (ma^2/\epsilon_h)^{1/2}= 3 -5$ ps. We use $\tau_L = 4.0 ps$ to convert the simulation times into real times. To estimate the timescale for mechanical unfolding dynamics, we use the Brownian dynamics algorithm (Ermack and McCammon, 1978; Klimov et al., 1998), for which the characteristic time for the overdamped motion is tH. We used $\zeta = 100 \tau^{-1}_L$ in the overdamped limit, and this value approximately corresponds to the friction constant for a molecule in water. All of the force simulations are performed at T = 300 K. For the integration time step, $h = 0.1\tau_L$, which implies that $10^6$ integration time steps correspond to 47 ms.\\

{\bf Dynamics of Rupture of Contacts: }
The time evolution of the rupture process is monitored by using the number of residues or nucleotide-dependent native contacts, $Q_i(t)$, that remain at t. We define $Q_i(t)=\sum^N_{j(|j-i|>2)}\Theta(R_C-r_{ij}(t))\Delta_{ij}$, where $R_C$ is the cut-off distance for native contacts, $r_{ij}(t)$ is the distance between the i-th and the j-th bead, and $\Delta_{ij} = 1$ for native contact (otherwise $\Delta_{ij} = 0$). If a certain subdomain of the molecule is disrupted and loses its contacts, then the extension of the molecule suddenly increases and the mechanical force exerted on the end of the molecule drops instantly. These molecular events are reflected as rips in the FEC. By comparing the time dependence of the force, $f(t)$, or the end-to-end distance, $R(t)$, with $Q_i(t)$ by using $t$ as a progressive variable to describe unfolding (see Figures 1C and 1D), we can unambiguously identify the structures involved in the dynamics of contact rupture.\\

{\bf Acknowledgments: }
We thank Prof. Matthias Rief and Hendrik Dietz for discussions on GFP. This work was supported in part by a grant from the National Science Foundation through grant number NSF-CHE-05-14056.
\\

{\bf REFERENCE}

Baldwin, R.L., and Rose, G.D. (1999). Is protein folding hierarchic? II. Folding intermediates and transition states. Trends Biochem. Sci. 24, 77-83.

Best, R.B., and Hummer, G. (2005). Comment on Force-clamp spectroscopy monitors the folding trajectories of a single protein. Science 308, 498b.

Brion, P., and Westhof, E. (1997). Hierarchy and dynamics of RNA folding. Annu. Rev. Biophys. Biomol. Struct. 26, 113-137.

Cecconi, C., Shank, E.A., Bustamante, C., and Marqusee, S. (2005). Direct observation of three-state folding of a single protein molecule. Science 309, 2057-2060.

Chen, S.J., and Dill, K.A. (2000). RNA folding energy landscapes. Proc. Natl. Acad. Sci. USA 97, 646-651.

Clementi, C., Nymeyer, H., and Onuchic, J.N. (2000). Topological and energetic factors: what determines the structural details of the transition state ensemble and en-route 
intermediates for protein folding? An investigation for small globular protein. J. Mol. Biol. 298, 937-953.

Das, R., Kwok, L., Millett, I., Bai, Y., Mills, T., Jacob, J., Maskel, G., Seifert, S., Mochrie, S., Thiyagarajan, P., et al. (2003). The fastest global events in RNA folding: electrostatic relaxation and tertiary collapse of the Tetrahymena ribozyme. J. Mol. Biol. 332, 311-319. 

Deras, M., Brenowitz, M., Ralston, C.Y., Chance, M.R., and Wood-son, S.A. (2000). Folding mechanism of the Tetrahymena ribozyme P4-P6 domain. Biochemistry 39, 10975-10985.

Derenyi, I., Bartolo, D., and Ajdari, A. (2004). Effects of intermediate bound states in dynamic force spectroscopy. Biophys. J. 86, 1263-1269.

Dietz, H., and Rief, M. (2004). Exploring the energy landscape of GFP by single-molecule mechanical experiments. Proc. Natl. Acad. Sci. USA 101, 16192-16197.

Enoki, S., Saeki, K., Maki, K., and Kuwajima, K. (2004). Acid denaturation and refolding of green fluorescence protein. Biochemistry 43, 14238-14248.

Ermack, D.L., and McCammon, J.A. (1978). Brownian dynamics with hydrodynamic interactions. J. Chem. Phys. 69, 1352-1369.

Fernandez, J.M., and Li, H. (2004). Force-clamp spectroscopy monitors the folding trajectory of a single protein. Science 303, 1674-1678.

Fukuda, H., Arai, M., and Kuwajima, K. (2000). Folding of green fluo-rescent protein and the cycle3 mutant. Biochemistry 39, 12025-12032.

Gao, M., Willmans, M., and Schulten, K. (2002). Steered molecular dynamics of titin domain unfolding. Biophys. J. 83, 3435-3445.

Gerland, U., Bundschuh, R., and Hwa, T. (2003). Mechanically probing the folding pathway of single RNA molecules. Biophys. J. 84, 2831-2840.

Hyeon, C., and Thirumalai, D. (2006). Forced-unfolding and force-quench refolding of RNA hairpins. Biophys. J. 90, 3410-3427.

Isralewitz, B., Gao, M., and Schulten, K. (2001). Steered molecular dynamics and mechanical functions of proteins. Curr. Opin. Struct. Biol. 11, 224-230.

Klimov, D.K., and Thirumalai, D. (2000a). Native topology determines force-induced unfolding pathways in globular proteins. Proc. Natl. Acad. Sci. USA 97, 7254-7259.

Klimov, D.K., and Thirumalai, D. (2000b). Mechanisms and kinetics of b-hairpin formation. Proc. Natl. Acad. Sci. USA 97, 2544-2549.

Klimov, D.K., Betancourt, M.R., and Thirumalai, D. (1998). Virtual atom representation of hydrogen bonds in minimal off-lattice models of a helices: effect on stability, cooperativity and kinetics. Fold. Des. 3, 481-498.

Laederach, A., Scherbakova, I., Liang, M.P., Brenowitz, M., and Altman, R.B. (2006). Local kinetic measures of macromolecular structure reveal partitioning among multiple parallel pathways from the earliest steps in the folding of a large RNA molecule. J. Mol. Biol. 358, 1179-1190.

Lehnert, V., Jaeger, L., Michel, F., and Westhof, E. (1996). New loop-loop tertiary interactions in self-splicing introns of subgroup IC and ID: a complete 3D model of the \emph{Tetrahymena thermophila} ribozyme. Chem. Biol. 3, 993-1009.

Li, M.S., Hu, C.K., Klimov, D.K., and Thirumalai, D. (2006). Multiple stepwise refolding of immunoglobulin I27 upon force quench depends on initial conditions. Proc. Natl. Acad. Sci. USA 103, 99-104.

Liphardt, J., Onoa, B., Smith, S.B., Tinoco, I., Jr., and Bustamante, C. (2001). Reversible unfolding of single RNA molecules by mechanical force. Science 292, 733-737.

Onoa, B., Dumont, S., Liphardt, J., Smith, S.B., Tinoco, I., Jr., and Bustamante, C. (2003). Identifying kinetic barriers to mechanical unfolding of the T. thermophila ribozyme. Science 299, 1892-1895.

Onuchic, J.N., and Wolynes, P.G. (2004). Theory of protein folding. Curr. Opin. Struct. Biol. 14, 70-75.

Pabon, G., and Amzel, L.M. (2006). Mechanism of titin folding bby force: insights from quasi-equilibrium molecular dynamics calculations. Biophys. J. 91, 467-472.

Rangan, P., Masquida, B., Westhof, E., and Woodson, S.A. (2003). Assembly of core helices and rapid tertiary folding of a small bacterial group I ribozyme. Proc. Natl. Acad. Sci. USA 100, 1574-1579.

Sali, A., Glaeser, R., Earnest, T., and Baumeister, W. (2003). From words to literature in structural proteomics. Nature 422, 216-225.

Scalvi, B., Sullivan, M., Chance, M.R., Brenowitz, M., and Woodson, S.A. (1998). RNA folding at millisecond interval by synchrotron hydroxyl radical footprinting. Science 279, 1940-1943.

Sosnick, T., and Pan, T. (2003). RNA folding: models and perspectives. Curr. Opin. Struct. Biol. 13, 309-316.

Takamoto, K., Das, R., He, Q., Doniach, S., Brenowitz, M., Herschlag, D., and Chance, M.R. (2004). Principles of RNA compaction: equilibrium folding pathway of the P4-P6 domain in monovalent cations. J. Mol. Biol. 343, 1196-1206.

Thirumalai, D., and Hyeon, C. (2005). RNA and protein folding: common themes and variations. Biochemistry 44, 4957-4970.

Thirumalai, D., and Woodson, S.A. (1996). Kinetics of protein and RNA folding. Acc. Chem. Res. 29, 433-439.

Thirumalai, D., Lee, N., Woodson, S.A., and Klimov, D.K. (2001). Early events in RNA folding. Annu. Rev. Phys. Chem. 52, 751-762. 

Treiber, D.K., and Williamson, J.R. (2001). Beyond kinetic traps in RNA folding. Curr. Opin. Struct. Biol. 11, 309-314.

Uchida, T., He, Q., Ralston, C.Y., Brenowitz, M., and Chance, M.R. (2002). Linkage of monovalent and divalent ion binding in the folding of the P4-P6 domain of the Tetrahymena ribozyme. Biochemistry 41, 5799-5806.

Veitshans, T., Klimov, D.K., and Thirumalai, D. (1997). Protein folding kinetics: timescales, pathways and energy landscape in terms of sequence-dependent properties. Fold. Des. 2, 1-22.

Zimmer, M. (2002). Green fluorescent protein (GFP): applications, structure, and related photophysical behavior. Chem. Rev. 102, 759-781.

\clearpage

\begin{table}
\begin{minipage}{\textwidth}
  \caption[Caption for LOF]%
  {Parameters for topology model of RNA and proteins.}
  \begin{tabular}{|c||c|c|}
  \hline
       &RNA & protein\\
  \hline
  $R_0$&2\AA&2\AA\ \\
  $k$& 20$kcal/(mol\cdot\AA^2)$ & 20$kcal/(mol\cdot\AA^2)$ \\
  $R_C$ & 14\AA\  & 8\AA\ \\
  $\epsilon_h$& 0.7$kcal/mol$& $(1-2)$ $kcal/mol$\\
  $\epsilon_l$& 1$kcal/mol$& 1$kcal/mol$\\
  $\sigma$ & 7\AA\  & 3.8\AA\ \\
  $\zeta$ &100$\tau_L^{-1}$&50$\tau_L^{-1}$\\
  $\tau_L$& 4$ps$ & 3$ps$\\
  \hline
  \end{tabular}
  \label{table:parameter}
\end{minipage}
\end{table}
\clearpage

\section{Figure Captions}

{{\bf Figure \ref{PNAS_FIG1} :}
Forced-Unfolding Dynamics of the \emph{Tetrahymena thermophila} Ribozyme
(A) Secondary structure of the \emph{Tetrahymena thermophila} ribozyme. Each subdomain is specified from P1 to P9. The four major tertiary interactions identified by the orange line are: (a) P2-P5c, (b) P2.1-P9.1a, (c) P5b-P6a, and (d) P5-P9.
(B) Superposition of 51 FECs for the T. thermophila ribozyme at r = 1.88 3104 pN/s, where $r_f= k v$ ($v = 5.4 \mu m/s$, $k = 3.5 pN/nm$). The value of $r_f$ is
about 3780 times greater than in the work by Onoa et al. (2003). The arrows indicate the rips.
(C and D) The number of rips varies between six and eight. The sharp peak preceding the first arrow (also in [C] and [D]) corresponds to unbinding of the extended 30 strand from the rest of the P4 subdomain. This peak is absent in the LOT experiments because the 30 end is shorter in the L-21 construct. (C) Superposition of FEC for 23 trajectories in which rupture begins with unraveling of the P9 helix. The arrows identify the position of the rips. The structures that unravel are explicitly indicated. The dynamics of disruption of individual contacts ($Q_i(t)$) (lower panel) for 1 of the 23 trajectories in which unfolding begins with P9 opening. The scale on the right in differing shades gives the number of contacts that survive at t. The structures in circles (a-d) are shown in (A), and the squares indicate interactions that stabilize the P3 pseudoknot. (D) Same as (C), except for this class of 28 trajectories, unfolding occurs by an alternate pathway in which the initial event is the opening of P2. The panel below gives $Q_i(t)$ for one of the trajectories.

{{\bf Figure \ref{PNAS_FIG2} :}
Snapshots of the Structures in the Unfolding of the T.
(A and B) The structures at various times along the pathways on the top and the bottom are obtained by the disappearance of the contacts $Q_i(t)$. In (A), the intermediates in the class of molecules in which P9 opens first is given; in (B), the structures from the alternate pathway in which P2 unravels initially are displayed. Along the pathway on the bottom, the P2.1-P9.1a tertiary contact is clearly observed in the structure at $t = 13.3 ms$. The structures in (A) and (B) are for the molecules whose contact rupture history is give in the two panels below Figures 1C and 1D, respectively.

{{\bf Figure \ref{PNAS_FIG3} :}
Loading Rate-Dependent Tension Propagation and Mechanical Unfolding of the Azoarcus Ribozyme (A) Secondary structure of the Azoarcus ribozyme.
(B) Force-extension curves of the Azoarcus ribozyme at three $r_f$s ($v = 43 \mu m/s$, $k_s = 28 pN/nm$ in red, $v = 12.9 \mu m/s$, $k_s = 28 pN/nm$ in green, and
$v = 5.4 \mu m/s$, $k_s = 3.5 pN/nm$ in blue).
(C) Contact rupture dynamics at three loading rates. The rips, resolved at the nucleotide level, are explicitly labeled.
(D) Topology of the Azoarcus ribozyme in the SOP representation. The first and the last alignment angles between the bond vectors and the force
direction are specified.
(E) Time evolutions of $\cos{\theta_i}$ (i = 1, 2, $\ldots$, N-1) at three loading rates are shown. The values of cosqi are color-coded as indicated on the scale shown on the right of the bottom panel.
(F) Comparisons of the time evolution of $\cos{\theta_i}$ (blue) and $\cos{\theta_{N-1}}$ (red) at three loading rates shows that the differences in the fc values at the opposite ends of the ribozyme are greater as $r_f$ increases.

{{\bf Figure \ref{PNAS_FIG4} :}
Force-Quench Refolding of the P4-P6 Domain (A) Refolding dynamics of the P4-P6 domain of the T. thermophila ribozyme upon force quench starting from a fully stretched state monitored by Q(t). The structure of the intermediate is explicitly shown.
(B) Dynamics of the end-to-end distance, $R(t)$ (black); the radius of gyration, $R (t)/R^N_G$ (green) 
($R^N_G = 2.98 nm$ for the native structure); and the root mean square deviation, $\Delta(t)$ (blue), with respect to the native state. In this trajectory, the helices form in $\sim 1.5 ms$, and the transition from a metastable to a native structure occurs in $\sim 4 ms$. The minimum $\Delta$ at long times is $6.4$ \AA.
(C) Dynamics of contact formation for the trajectory in (A). The early stage of folding is related to the hairpin loop formation (enclosed in the red box) that is separately shown on the right. Zipping of the secondary structure propagates from around P5b, P5c, and P6b. Formations of the tertiary contacts (w6 ms) in P5a/P5c-P4 and in P5b-P6a are shown in circles.

{{\bf Figure \ref{PNAS_FIG5} :}
Dominant Force-Induced Unfolding Pathway of GFP
(A) Native structure of GFP (chain A from PDB code: 1gfl; 230 residues). (B) Two-dimensional connectivity map of GFP (top view of (A)).
(C) Force-extension curve (FEC) at the pulling speed of $v = 2.5 \mu m/s$ and $k_s = 35 pN/nm$. The secondary structural elements that unravel at the rips
are explicitly indicated. The purple arrow indicates that strands $\beta1-\beta3$ transiently form before further unfolding.
(D) Dynamics of rupture of contacts formed by each residue, $Q_i(t)$, corresponding to the FEC in (C). The scale on the right in different shades gives the number of contacts that remain at time t.
(E) The structures involved in the unfolding pathway labeled (i)-(iv) are shown by using the color code: $\beta2$, purple; $\beta3$, purple; $\beta4$, green; $\beta5$, green; $\beta6$, green; $\beta7$, red; $\beta8$, red; $\beta9$, red; $\beta10$, cyan; and $\beta11$, yellow. The N-terminal helix, H1, is in pink. The secondary structure elements that unfold together upon application of force at the C-terminal end are shown in the same color.

{{\bf Figure \ref{PNAS_FIG6} :} 
Force-Induced Unfolding of GFP: Minor Pathway
(A) FEC for GFP unfolding by an alternate pathway. The secondary structural elements that are disrupted in the early stages are explicitly indicated.
(B) Structures that are populated in the transition to the stretched states for the trajectory in (A).

{{\bf Figure \ref{PNAS_FIG7} :} 
Refolding of GFP upon Force Quench (A) Dynamics of approach to the native state upon force quench is monitored by $Q(t)$. The inset shows the time dependence of $R(t)$,
$R_G(t)$, and $\Delta(t)$.
(B) Time-dependent formation of the native contacts at the residue level during refolding from stretched GFP.

\newpage
\begin{figure}[ht]
\includegraphics[width=7.00in]{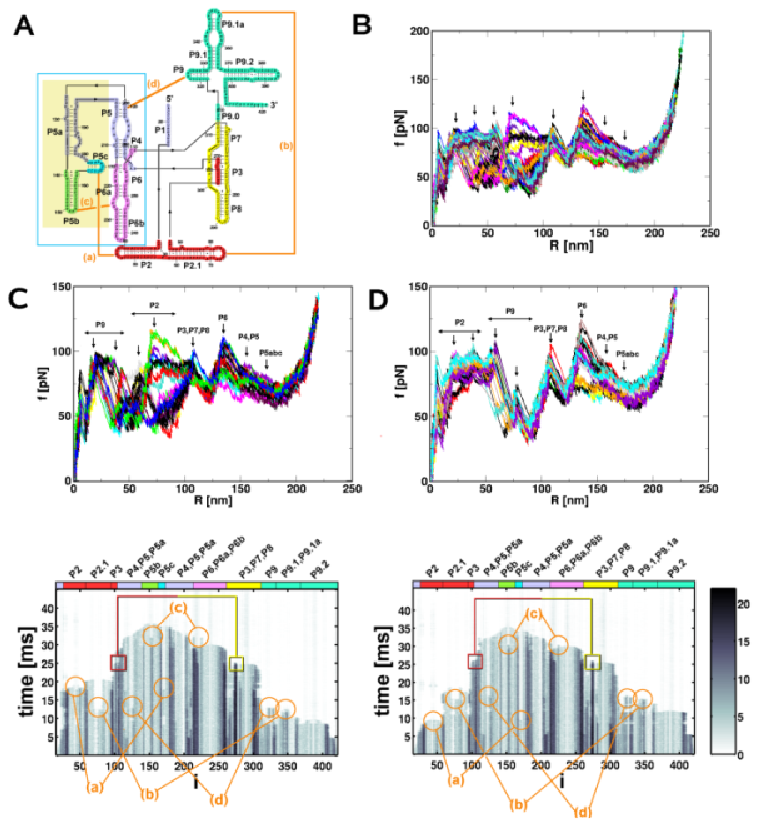}
\caption{\label{PNAS_FIG1}}
\end{figure}
\begin{figure}
\centering
\includegraphics[width=7.00in]{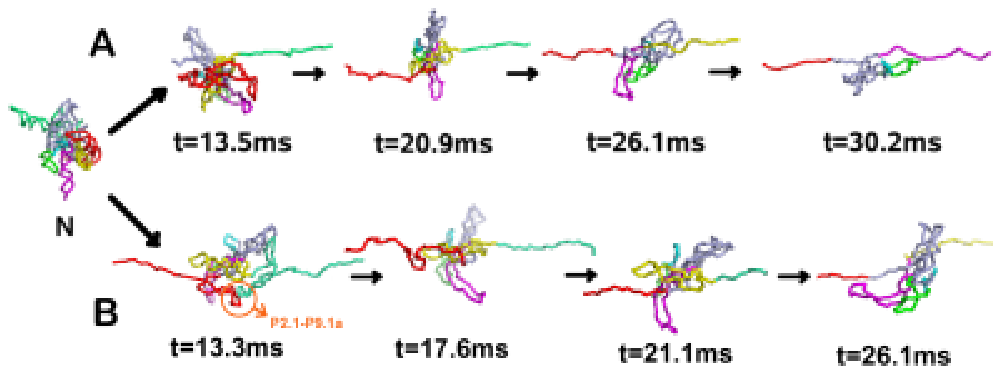}
\caption{\label{PNAS_FIG2}}
\end{figure}
\begin{figure}
\centering
\includegraphics[width=5.00in]{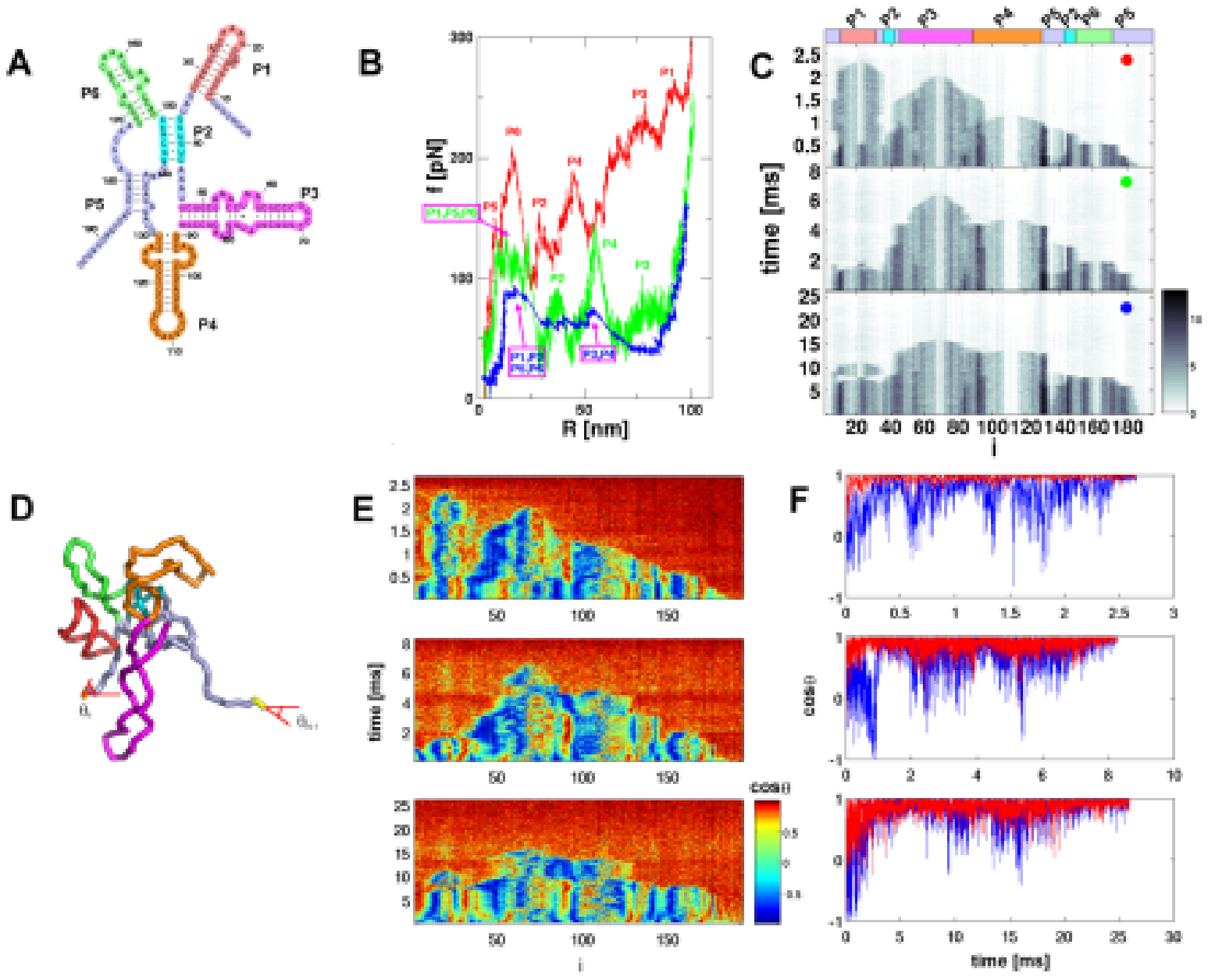}
\caption{\label{PNAS_FIG3}}
\end{figure}
\newpage
\begin{figure}
\centering
\includegraphics[width=4.0in]{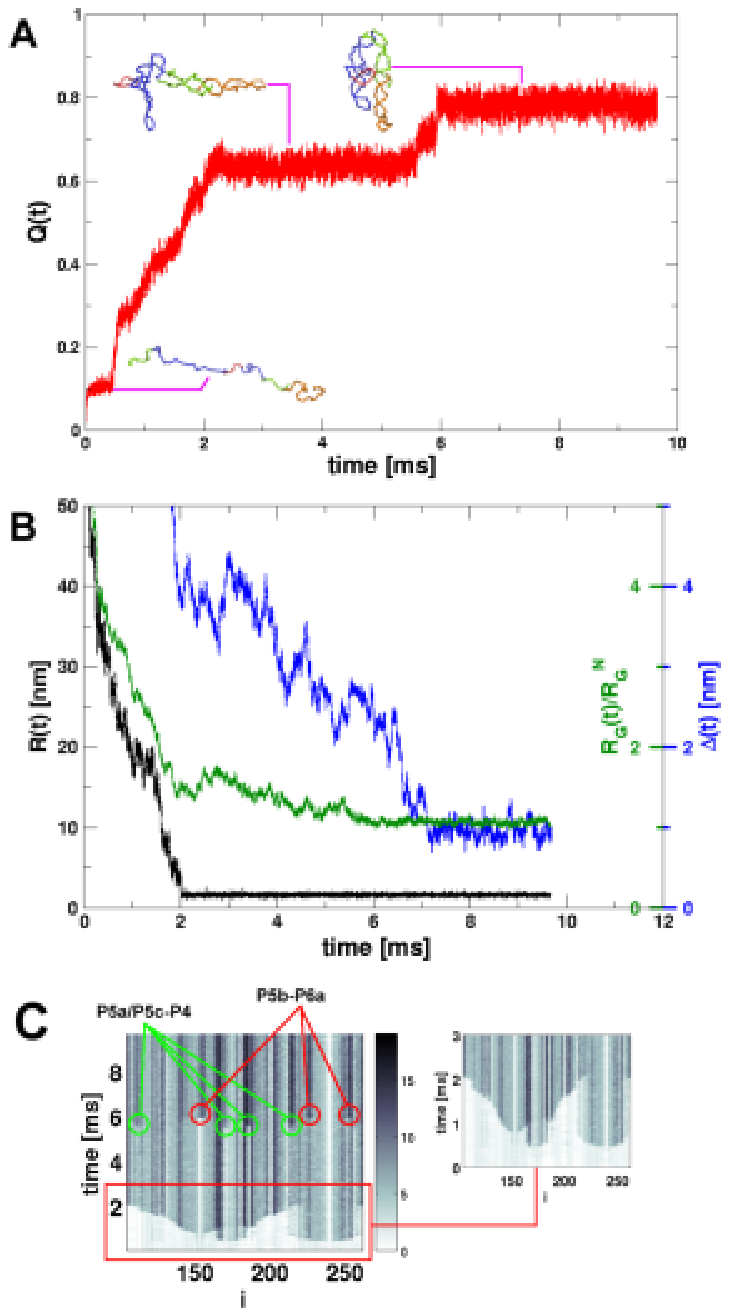}
\caption{\label{PNAS_FIG4}}
\end{figure}
\begin{figure}
\centering
\includegraphics[width=5.5in]{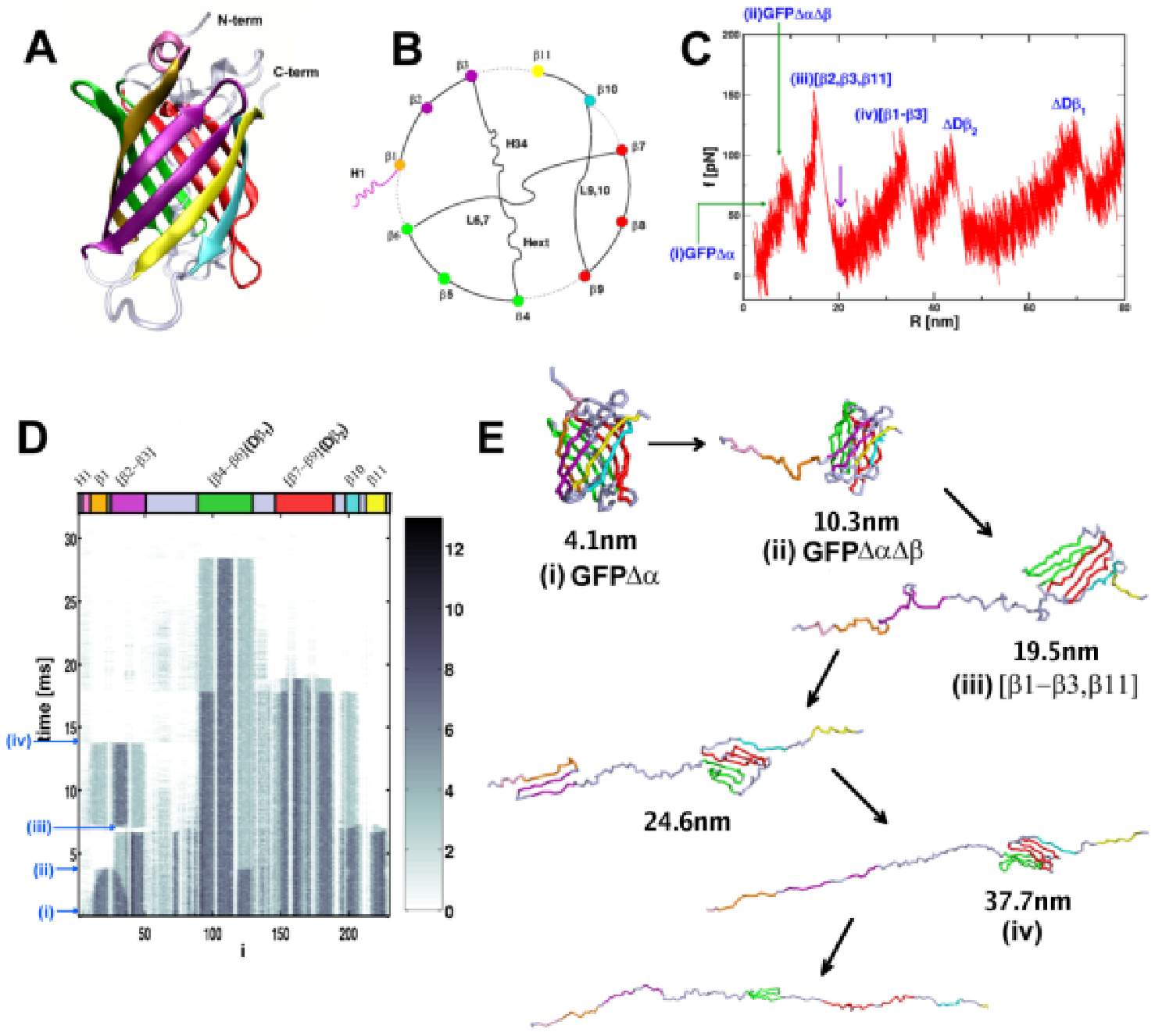}
\caption{\label{PNAS_FIG5}}
\end{figure}
\begin{figure}
\centering
\includegraphics[width=6.0in]{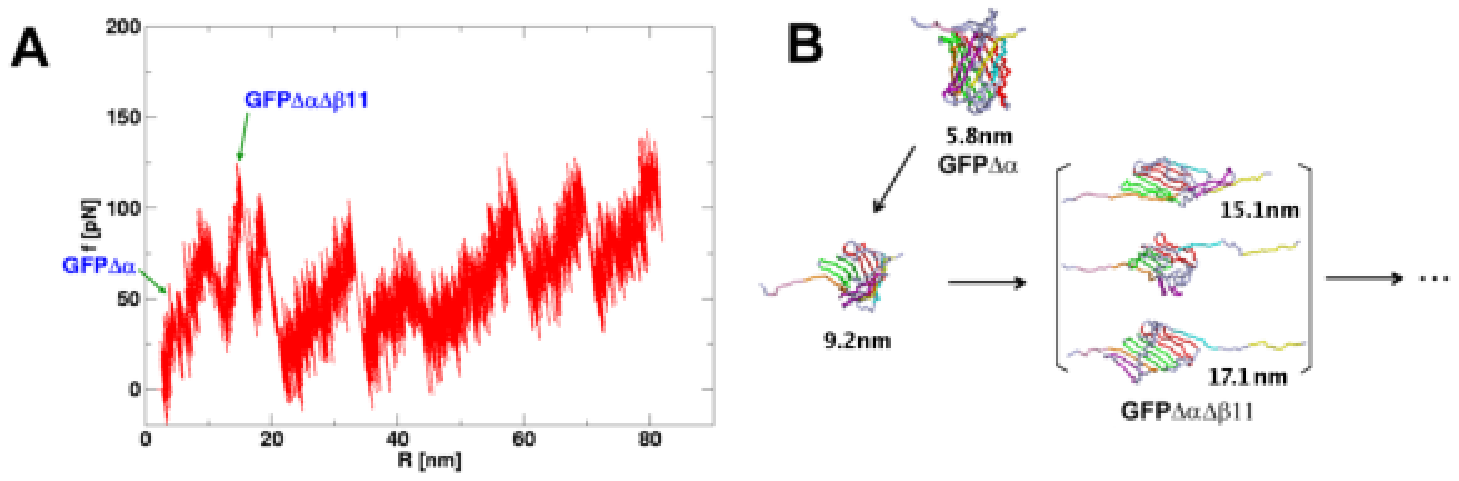}
\caption{\label{PNAS_FIG6}}
\end{figure}
\begin{figure}
\centering
\includegraphics[width=4.0in]{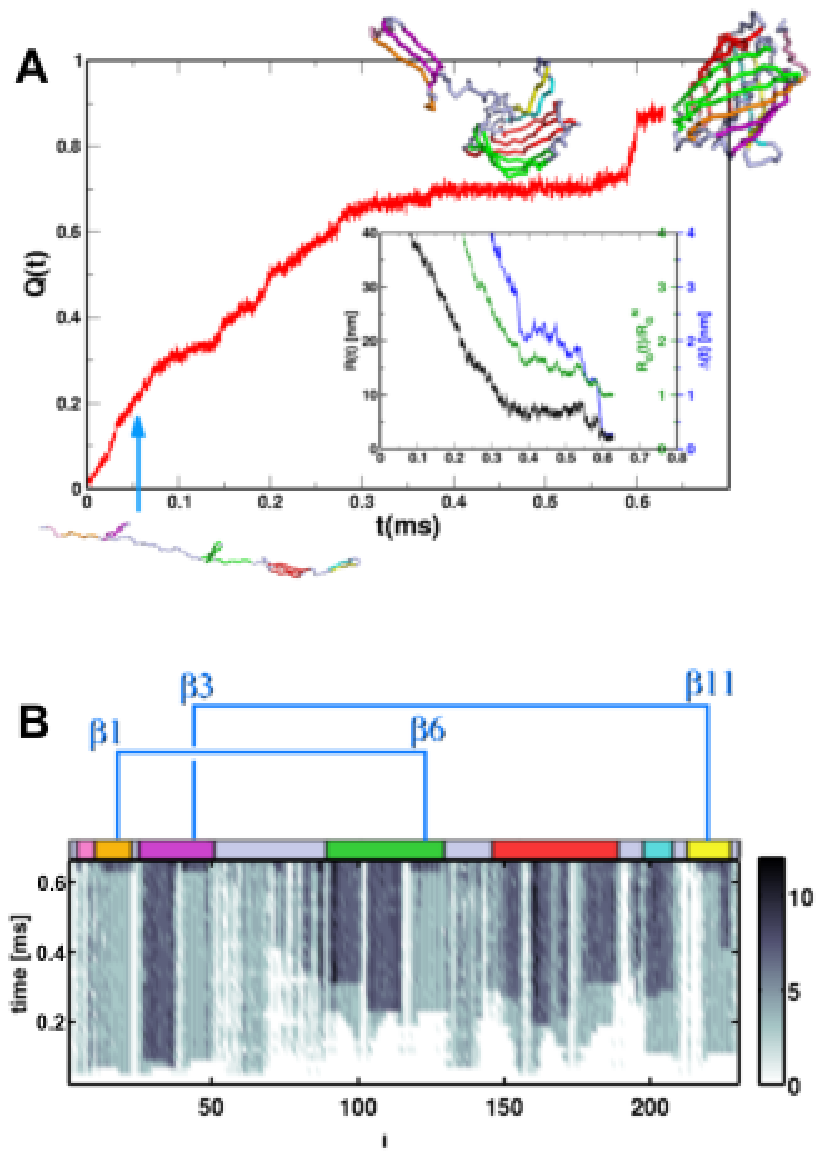}
\caption{\label{PNAS_FIG7}}
\end{figure}
\end{document}